\documentclass[aps,pra,twocolumn,nofootinbib,groupedaddress,floatfix,showpacs]{revtex4}
\usepackage{graphicx}
\usepackage{amssymb}
\usepackage{amsmath}
\usepackage{amsfonts}
\usepackage{multirow}
\usepackage{longtable}

\begin{document}

\title{
Electron Chirality in Amino Acid Molecules}

\author{Masato Senami}
\email{senami@me.kyoto-u.ac.jp}
\affiliation{Department of Micro Engineering, Kyoto University, Kyoto 615-8540, Japan}
\author{Tomoki Shimizu}
\affiliation{Department of Micro Engineering, Kyoto University, Kyoto 615-8540, Japan}

\date{\today}

\begin{abstract}
We evaluated the total electron chirality in alanine, serine, and valine,
which are molecules that have chiral structures.
Previously, it has been considered that 
the total electron chirality of molecules composed of only light elements
cannot be computed within usual computational conditions of relativistic four-component wave functions.
In this work, it is shown that the total electron chirality can be calculated 
if some diffuse functions are added to Gaussian basis sets.
This is demonstrated for the H$_2$O$_2$ molecule.
By adding diffuse Gaussian functions to basis sets,
the total electron chirality of L-alanine, L-serine, and L-valine are evaluated.
It is also shown that 
the total electron chirality is derived
by the cancellation between large contributions from each orbital,
and the total electron chirality in excited and ionized states 
is expected to be much larger than that of the ground state.
\end{abstract}

\pacs{31.30.J-, 31.30.jg, 95.30.Ft}

\maketitle
\section{Introduction}

Electrons are chiral objects from the viewpoints of the fundamental physics.
The weak interaction, which is one of the four fundamental interactions in particle physics,
produces larger reaction rates for left-handed particles than for right-handed particles \cite{textbook1}.
Electron is a point particle,
and it is considered to be spherically symmetric.
However, electron has velocity and spin angular momentum,
and the chirality of the electron can be defined
on the basis of the relativistic form of these quantities.

In our previous works \cite{Senami:2019,Inada:2018},
it has been established that 
electron chirality is biased toward a particular form, L- or D-configuration,
in chiral molecules owing to spin-orbit interactions
and the electron chirality in L-configuration molecules 
is opposite of that in D-configuration molecules.
As mentioned above, the interaction of electrons is dependent on electron chirality.
Hence, L- and D-enantiomers exhibit different interaction properties
in the weak interaction.

Electron chirality in chiral molecules (ECCM) may be related to some unsolved problems.
One example is the chiral-induced spin selectivity (CISS) effect
(e.g., \cite{CISS}),
which is the polarization of the electron spin through chiral organic molecules.
It is unclear how this polarization is generated through chiral molecules.
For example, some researchers consider that spin-orbit interactions induce this effect;
however a two orders of magnitude larger spin-orbit interaction than that of a carbon atom is needed.
It has been shown that 
electron chirality is one of the origins of the torque for the electron spin
in the quantum field theoretical picture \cite{spintorque,Text:QED}.
Therefore, ECCM may be a good solution for CISS.

Another example is the homochirality in nature.
In our previous paper \cite{Senami:2019},
we pointed out that ECCM may be the solution of homochirality.
On Earth, amino acids and sugars are extremely biased in chirality.
Almost all amino acids have L-configuration,
while sugars predominantly exist in D-configuration.
It is believed that a small source of enantiomeric asymmetry is produced;
then, this enantiomeric excess is enhanced to large imbalance through chiral amplification processes.
It is unclear how this initial bias is generated during the evolution of the universe or Earth.
Thus, far several scenarios for the production of this initial imbalance
have been proposed \cite{text:homochirality}.
For example, this imbalance has been proposed to be produced by 
circularly polarized light from pulsars,
the parity-violating energy difference between enantiomers by the weak interaction,
or left-handed electrons by beta decay of nuclei 
(the Vester-Ulbricht hypothesis \cite{Vester:1959}).
In our previous paper, we added ECCM to this list \cite{Senami:2019}.
One form of an enantiomeric pair has different reaction rates of the weak interaction owing to ECCM
because left-handed electrons interact more strongly than right-handed ones by the weak interaction.
Thus, one-configuration of an enantiomeric pair is more fragile than the rest of the pair
by the interactions with particles such as cosmic rays and neutrinos produced by nuclear fusion in stars.
This gives inevitably nonzero contribution to the difference in the number of enantiomers.

These two examples are driven by molecules consisting of only light elements
up to the second row in the periodic table.
In our previous work \cite{Senami:2019}, 
the reliable result of the total chirality of H$_2$O$_2$,
which is one of the simplest examples of chiral molecules consisting of only light elements,
could not be obtained by ordinary four-component relativistic quantum structure computations.
To study these two examples,
a reliable computational method is mandatory.

In this study,
first, we establish a computational method of the total chirality of 
chiral molecules consisting of only light elements.
For this purpose, an H$_2$O$_2$ molecule is used,
and computations are determined to be reliable 
if the dependence of the total chirality on the dihedral angle 
is the same as that of other H$_2$X$_2$ molecules.
Because the parity-violating energy difference,
which primarily depends on the electron chirality at the positions of nuclei,
was accurately calculated \cite{Inada:2018},
computed wave functions around nuclei are sufficiently accurate.
Thus, the addition of diffuse functions to basis sets is speculated to be effective.
In this paper, we will confirm this speculation.
Second, the total electron chirality of amino acid molecules is calculated 
for alanine, serine, and valine using basis sets containing additional diffuse functions.
For our scenario of the generation of homochirality,
L-enantiomers of amino acid molecules should have a smaller ratio of left-handed electrons,
which corresponds to the positive value of chirality.

This paper is organized as follows.
In the next section, 
the definition of the total electron chirality 
and our computational method are explained.
In Sec. \ref{sec:results}, our results are presented.
The dependence of the total electron chirality of H$_2$O$_2$ on the dihedral angle 
is calculated for various basis sets,
and it is established that the addition of diffuse functions is effective for 
the computation of the total electron chirality of chiral molecules consisting of only light elements.
Then, the total electron chirality of L-alanine, L-serine, and L-valine is evaluated;
it is determined that all of the above-mentioned molecules,
and they all exhibit positive chirality values.
The total electron chirality is shown to be the result of
the cancellation between large contributions from each orbital.
The last section is devoted to our summary.

\section{Theory and Computational Details}

Electron chirality is classified into right- or left-handed electrons.
The right- and left-handed projection operators of the electron chirality are defined by:
\begin{align}
P_R = \frac{1 + \gamma_5}{2},
\ \ \ 
P_L = \frac{1 - \gamma_5}{2},
\end{align}
where $ \gamma_5= i \gamma^0 \gamma^1 \gamma^2 \gamma^3$
and $\gamma^\mu~(\mu=0-3)$ are the gamma matrices.
In another expression,
\begin{align}
\gamma_5 = \frac{1}{3} \sum_{i=1}^3 \alpha^i \Sigma^i 
\end{align}
where $\alpha^i = \gamma^0 \gamma^i$ and
$ \Sigma^i $ is the $4 \times 4$ extension of the Pauli matrix.
In this form, chirality is realized as the relativistic counterpart of helicity,
and the operator, $\gamma_5$, exhibits the characteristic of the parity odd \cite{textbook1, textbook2}.
The right- and left-handed electrons are described by the following projection operators:
\begin{align}
\psi_R (\vec r) = P_R \psi (\vec r), 
\ \ \ 
\psi_L (\vec r) = P_L  \psi (\vec r),
\end{align}
where $\psi$ is the four-component wave function of the electron.
The difference in the density between the right-handed and left-handed electrons
is described by the chirality density as follows:
\begin{align}
\psi^\dagger (\vec r) \gamma_5 \psi (\vec r)
=
\psi_R^\dagger (\vec r) \psi_R (\vec r) 
- \psi_L^\dagger (\vec r) \psi_L (\vec r).
\end{align}
In the computational chemistry field,
chirality density was first taken into account in Ref.~\cite{Hegstrom:1985}
after many textbooks on relativistic quantum theory explained electron chirality.
The total electron chirality is obtained by 
integrating the chirality density over an entire molecule,
\begin{align}
\int d^3 \vec r \psi^\dagger (\vec r) \gamma_5 \psi (\vec r) .
\end{align}
The total electron chirality is a different notion from the parity-violating energy difference
(e.g., see \cite{Bast:2011}).
The parity-violating energy difference depends only on 
the electron chirality at the positions of nuclei in a molecule,
while the total electron chirality is obtained by integrating the entire molecule.

Electronic structure should be computed in a fully relativistic form,
i.e., a four-component wave function.
This computation was carried out using the DIRAC17 program package \cite{DIRAC17}.
We adopted the relativistic Hartree-Fock method with the Dirac-Coulomb Hamiltonian.
For the derived wave functions,
the total electron chirality was calculated by the QEDynamics program package \cite{QEDynamics}.
For basis sets, we chose
dyall.ae3z, dyall.ae4z, cc-pV5Z, and cc-pV6Z
\cite{Dyallbasis, ccpv5z, aug-ccpv6z}.
For all basis sets, uncontracted forms were used.
The number of Gaussian functions of cc-pV5Z (cc-pV6Z) in the uncontracted form
is approximately similar to that of dyall.ae3z (dyall.ae4z).
Correlation consistent basis sets have more functions
for higher angular momentum functions, such as g or h.
However, correlation consistent basis sets
are not constructed with relativistic effects.

Moreover, diffuse functions are added to these basis sets.
For Dyall basis sets (dyall.ae3z and dyall.ae4z),
nomenclature, such as q-aug-dyall.ae3z, is adopted,
and diffuse functions are added as follows.
For each angular momentum,
the exponent factor ratio is determined
by the ratio of the exponents between the two most diffuse exponent basis functions.
If only one function exists in the angular momentum,
the ratio is chosen to be 1/3.5.
Using this ratio, new exponent functions are repeatedly added $n$ times,
where $n = $ 2, 3, 4 for $X$=d, t, q in $X$-aug-dyall.ae$m$z basis sets ($m =$ 3, 4).
If atomic species are specified as a prefix of a basis set name $(Atom)$-$X$-aug-dyall.ae$m$z,
diffuse functions are added only to the basis functions of specified atoms.
For correlation-consistent basis sets,
ordinary augmented basis sets are adopted, i.e., aug-cc-pV5Z, and aug-cc-pV6Z
\cite{aug-ccpv6z,aug-ccpv5z},
which add one small exponent function to each angular momentum.

 
 




\begin{figure}[p]
 \begin{center}
\includegraphics[width=5.5cm]{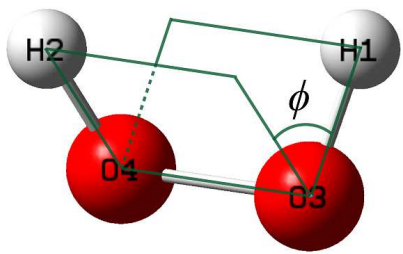} 
\put(-105,-10){\large (a) $\rm H_2 X_2$ }
\\
\vspace{4mm}
\includegraphics[width=5.4cm]{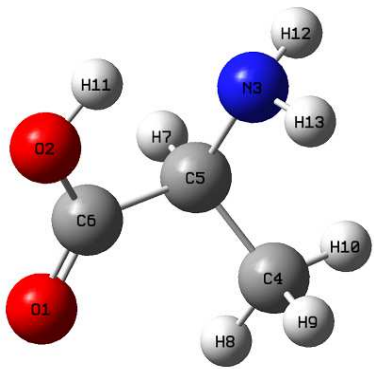} 
\put(-110,-10){\large (b) Alanine }
\\
\vspace{4mm}
\includegraphics[width=5.4cm]{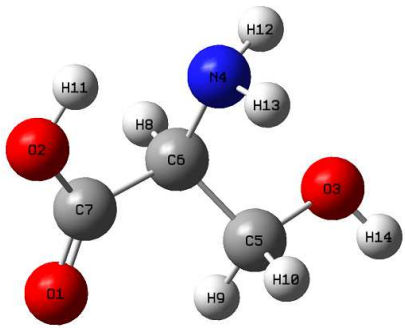} 
\put(-110,-10){\large (c) Serine }
\\
\vspace{4mm}
\includegraphics[width=5.4cm]{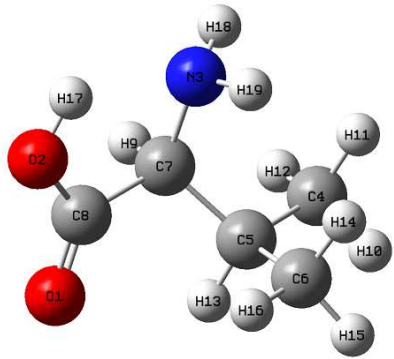} 
\put(-110,-10){\large (d) Valine }
\end{center} 
 \caption{Structures of (a)$\rm H_2 X_2$, (b) alanine, (c) serine, and (d) valine
molecules and the definition of the dihedral angle $\phi$ of $\rm H_2 X_2$.}
 \label{fig:models}
\end{figure}

In our computations, 
H$_2$O$_2$, H$_2$Te$_2$, alanine, serine, and valine were calculated.
The geometrical structure of H$_2$X$_2$ (X=O, Te) was set to be the same 
as that in our previous work \cite{Senami:2019}.
For alanine, serine, and valine molecules,
the geometrical optimization was performed with the dyall.ae2z basis set \cite{Dyallbasis}.
The structure of the molecules is shown in Fig.~\ref{fig:models}.
The dihedral angle $\phi$ of H$_2$X$_2$ is defined in this figure.
For the study of varying $\phi$, 
only $\phi$ is varied from this optimized geometry,
while the other parameters, such as internuclear lengths, are fixed.

In the following, we adopted the atomic unit unless stated otherwise.

\section{Results}
\label{sec:results}

\subsection{$\rm H_2 O_2$}
\label{sec:H2O2}

\begin{figure}[t]
 \centering
\includegraphics[width=0.85\linewidth]{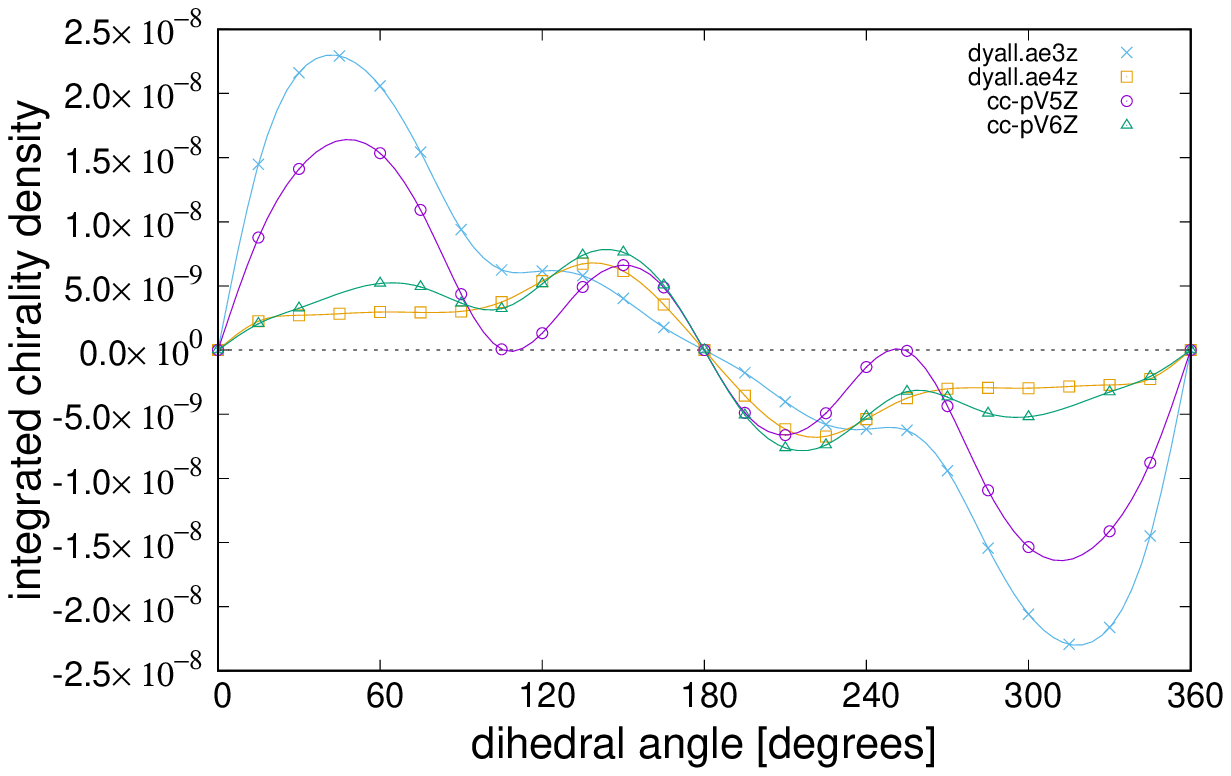} \\
$ $\put(0,10){\large $\rm H_2 O_2$ }
\\
\includegraphics[width=0.85\linewidth]{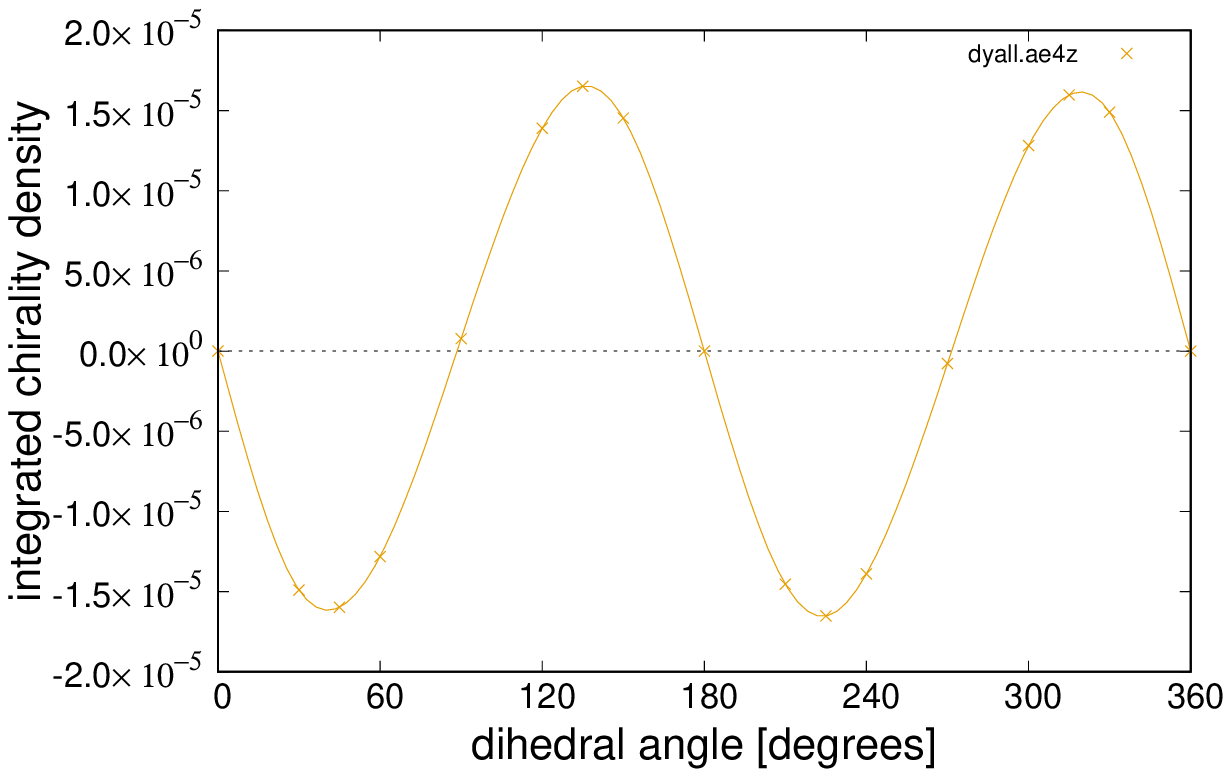} \\
$ $\put(0,0){\large $\rm H_2 Te_2$ }
\\
\caption{
Total chirality of $\rm H_2O_2$ and $\rm H_2Te_2$ as a function of the dihedral angle, $ \phi $.
}
\label{fig:chirality_wo}
\end{figure}

Figure \ref{fig:chirality_wo} shows 
the total electron chirality of $\rm H_2O_2$ and $\rm H_2 Te_2$
for the basis sets without additional diffuse functions. 
The dependence of the total chirality of $\rm H_2O_2$ on the dihedral angle, $\phi$,
should exhibit the same curve as that of $\rm H_2 Te_2$.
This occurs because the molecule structure governs the total chirality,
as reported in our previous works \cite{Inada:2018,Senami:2019},
where $\rm H_2 Se_2$, $\rm H_2 S_2$, and $\rm H_2 Te_2$ exhibited
the same dependence on $\phi$.
However, the result for $\rm H_2O_2$ shows a different dependence on $\phi$ from that of $\rm H_2 Te_2$.
Moreover, the curves of $\rm H_2O_2$ differ from each other.
This indicates that this result does not have a reliable accuracy.

\begin{figure}[t]
	\centering
\includegraphics[width=0.9\linewidth]{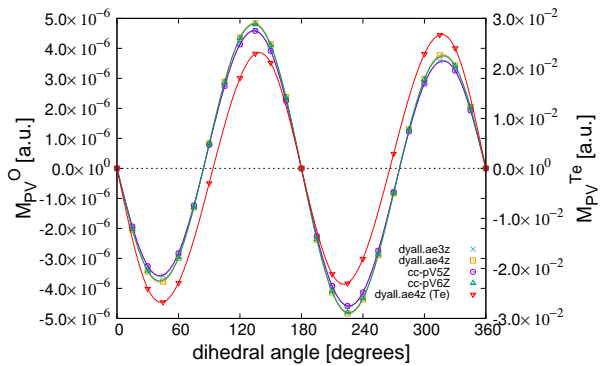} \\
	\caption{Electron chirality densities at the positions of O and Te nuclei
in $\rm H_2O_2$ and $\rm H_2Te_2$
		 as a function of $ \phi $.
	}
	\label{fig:gamma5_X}
\end{figure}

For the total electron chirality, the accuracy was not sufficient;
however, these computations were sufficiently accurate to represent the parity-violating energy difference.
In our previous study \cite{Inada:2018,Senami:2019},
the parity-violating energy difference of $\rm H_2O_2$
exhibited the same dependence as that of other $\rm H_2X_2$ molecules on $\phi$.
The parity-violating energy difference of $\rm H_2X_2$ molecules is predominantly determined by 
the electron chirality density at the X nucleus position.
Hence, our wave functions are sufficiently accurate around the nuclei,
and the accuracy is not sufficient in regions away from nuclei. 
To confirm that the wave functions in this study were accurate around O nuclei,
the electron chirality densities at O and Te nuclei, $M_{\rm PV}^{\rm X}$,
are shown in Fig.~\ref{fig:gamma5_X}.
For all basis sets, the curves at the O nucleus are consistent with that at the Te nucleus.
Therefore, our wave functions are confirmed to be sufficiently accurate around the nuclei.
It is speculated that wave functions should be improved in regions away from nuclei.
Thus, we add additional diffuse functions to basis sets.

\begin{figure}[t]
\centering
\includegraphics[width=0.85\linewidth]{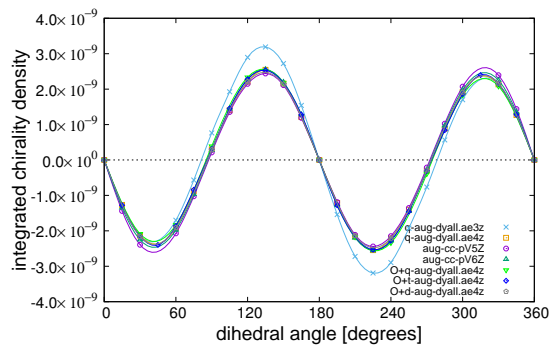} \\
\caption{Total electron chirality of $\rm H_2O_2$ as a function of the dihedral angle, $ \phi $,
with additional diffuse functions.
}
\label{fig:chirality}
\end{figure}

Figure \ref{fig:chirality} shows
the total electron chirality of $\rm H_2O_2$ as a function of the dihedral angle,
for the basis sets with additional diffuse functions.
It is observed that all results show the same dependence on $\phi$.
Specifically, the results of dyall.ae4z with additional diffuse functions are well-converged,
and sufficient accuracy is attained.
The number of additional diffuse functions for
Dyall basis sets is much larger than that for correlation-consistent ones.
As observed from the results of correlation-consistent ones,
many additional diffuse functions are not required,
i.e., only ordinary augmentation or twice of it is sufficient to 
evaluate the total electron chirality.
The results of cc-pV5Z and cc-pV6Z are very similar to those of dyall.ae3z and dyall.ae4z, respectively.
Hence, the relativistic effect in basis sets is seen to be less important than other relativistic effects.
The spin-orbit interaction of oxygen atoms is very small compared to S, Se, and Te atoms,
whose total electron chirality was studied in our previous works \cite{Inada:2018,Senami:2019}.
Accordingly, the accurate inclusion of the spin-orbit interaction was challenging,
and we accurately calculated it with additional diffuse functions.

\begin{table}[t]
  \centering
  \caption{Contribution of each orbital to the total electron chirality of H$_2$O$_2$ at $\phi =45^\circ$
for the result of q-aug-dyall.ae3z.
$E_i$ is the orbital energy and $\psi_i^\dagger (\vec r_O ) \gamma_{5} \psi_i (\vec r_O )$
is the electron chirality at the position of an oxygen nucleus.
The value of the orbital contribution to the electron chirality is shown for one electron in Kramers pair,
while both values in pairs are summed for the total.
}
  \begin{tabular}{c|rrr} \hline\hline
    No.   & $E_i$ [a.u.] & $ \psi_i^\dagger (\vec r_O ) \gamma_{5} \psi_i (\vec r_O )$
 & $\int d^3 x \psi_i^\dagger  \gamma_{5} \psi_i$ \\ \hline
    1     & ~ $-20.643$    & ~ $-3.688\times10^{-8}$  & ~ $-1.634\times10^{-9}$ \\
    2     & $-20.643$    & $-3.276\times10^{-8}$  &  $-1.441\times10^{-9}$ \\
    3     & $-1.517$     & $-3.307\times10^{-7}$  &  $ 1.540\times10^{-8}$ \\
    4     & $-1.214$     & $ 1.168\times10^{-7}$  &  $ 1.196\times10^{-8}$ \\
    5     & $-0.784$     & $-7.179\times10^{-6}$  &  $-1.649\times10^{-7}$ \\
    6     & $-0.674$     & $ 2.425\times10^{-5}$  &  $ 8.205\times10^{-7}$ \\
    7     & $-0.623$     & $-1.817\times10^{-5}$  &  $-7.642\times10^{-7}$ \\
    8     & $-0.548$     & $ 4.486\times10^{-6}$  &  $-7.593\times10^{-8}$ \\
    9     & $-0.463$     & $-4.902\times10^{-6}$  &  $ 1.591\times10^{-7}$ \\ \hline
    Total &              & $-3.589\times10^{-6}$  &  $-2.310\times10^{-9}$ \\ \hline
  \end{tabular}
  \label{Table:H2O2_orbital}
\end{table}

In our previous work \cite{Senami:2019},
we have reported that the total electron chirality of H$_2$Te$_2$ was derived 
as the result of the cancellation between large contributions from each orbital.
Actually, many higher energy orbitals, such as the highest occupied molecular orbital (HOMO),
have larger contribution than the total electron chirality.
Hence, it was speculated that 
if chiral molecules were excited or ionized,
this cancellation was broken and the chiral molecules had much larger total electron chirality.
In the work, we confirmed this speculation with imaginary H$_2$Te$_2^{2+}$.
To confirm that this speculation is true even for H$_2$O$_2$,
which is a molecule composed of only light elements
and has small spin-orbit interaction,
each orbital contribution to the total electron chirality of H$_2$O$_2$ is studied.
In Table \ref{Table:H2O2_orbital},
the contribution to the total electron chirality, $\int d^3 x \psi_i^\dagger  \gamma_{5} \psi_i$,
and that to the electron chirality at the position of an O nucleus,
$ \psi_i^\dagger (\vec r_O ) \gamma_{5} \psi_i (\vec r_O )$,
are shown for the result of q-aug-dyall.ae3z of the $\rm H_2 O_2$ molecule at $\phi = 45^\circ$,
where $ \psi_i $ is the $i$-th orbital.
The total electron chirality is $-2.310\times10^{-9}$.
Contributions from each orbital except for the lowest two orbitals
are considerably larger than even the total value.
Therefore, even for the molecules of only light elements,
excitation or ionization is considered to enhance 
the total electron chirality.
We will confirm this enhancement with configuration interaction computations in our next paper.
Moreover, this property is also observed for the electron chirality at the position of an O nucleus.
The parity-violating energy difference,
which is dominantly determined by the electron chirality at the position of an O nucleus,
is speculated to be enhanced by excitation or ionization.
The cancellation is not stronger than the total electron chirality,
and the enhancement of the parity-violating energy difference 
is not as large as the total electron chirality.
These features are the same as those of H$_2$Te$_2$.

\subsection{Amino Acids}
\label{sec:AA}

\begin{table*}[t]
 \caption{
Total electron chirality and the parity violating energy of L-amino acids.
}
  \begin{tabular}{c|ccc} \hline\hline
    molecule   &  basis set        & $\int d^3 x \psi^\dagger  \gamma_{5} \psi$ & $E_{\rm{PV}}$ [a.u.]   \\
 \hline
    L-Alanine  
& dyall.ae3z       & $-1.062\times10^{-9}$                & $-3.915\times10^{-20}$  \\ 
& ONC+t-dyall.ae3z & $1.708\times10^{-11}$                     & $-3.944\times10^{-20}$  \\ 
& ONC+q-dyall.ae3z & $2.954\times10^{-11}$                     & $-3.944\times10^{-20}$  \\ 
& dyall.ae4z       & $6.849\times10^{-10}$                     & $-4.137\times10^{-20}$  \\
 \hline
    L-Serine 
& dyall.ae3z       & $6.166\times10^{-9}$                      & $-4.369\times10^{-20}$  \\ 
& ONC+t-dyall.ae3z & $2.153\times10^{-10}$                     & $-4.407\times10^{-20}$  \\ 
& ONC+q-dyall.ae3z & $2.181\times10^{-10}$                     & $-4.407\times10^{-20}$  \\ 
& dyall.ae4z       & $2.417\times10^{-9}$                      & $-4.616\times10^{-20}$  \\
 \hline
    L-Valine
& dyall.ae3z       & $-5.213\times10^{-10}$                    & $-1.580\times10^{-20}$  \\ 
& ONC+t-dyall.ae3z & $1.887\times10^{-11}$                     & $-1.600\times10^{-20}$  \\ 
& ONC+q-dyall.ae3z & $2.952\times10^{-11}$     &               $-1.598\times10^{-20}$ \\ 
& dyall.ae4z       & $9.818\times10^{-11}$                     & $-1.681\times10^{-20}$  \\
 \hline\hline
  \end{tabular}
 \label{table:AAChirality}
\end{table*}

In the previous subsection,
we showed that the total electron chirality of light element molecules 
can be computed by extending basis sets with additional diffuse functions.
In this subsection, the results for amino acids are shown.

In Table \ref{table:AAChirality}, the total electron chirality of 
L-alanine, L-serine, and L-valine is shown for several basis sets.
Moreover, the parity-violating energy is also shown as a references.
The parity violating energy is calculated by the following equation,
\begin{align}
E_{\rm PV} 
=
\frac{G_F}{2 \sqrt{2}} \sum_n Q_W^n \psi^\dagger (\vec r_n) \gamma_5 \psi (\vec r_n),
\end{align}
where $G_F / (\hbar c)^3 = 1.166 \times 10^{-5} $ GeV$^{-2}$\cite{PDG}
and $\vec r_n$ is the position of a nucleus, $n$.
$Q_W^n = Z^n ( 1 - 4 \sin^2 \theta_W ) - N^n $ is the weak charge of a nucleus, $n$,
where $Z^n$ and $N^n$ are the number of protons and neutrons in the nucleus, $n$,
respectively, and $\theta_W$ is the weak-mixing angle,
whose value is determined by $\sin^2 \theta_W = 0.2312$ \cite{PDG}.
The parity-violating energy is almost independent of the choice of basis sets,
and the difference between dyall.ae3z and dyall.ae4z is within 10\%.
Moreover, our results are consistent with the reported value \cite{E_PV}.
However, the total electron chirality is completely different 
between dyall.ae3z and dyall.ae4z.
This is the same situation as that for H$_2$O$_2$.
The wave functions around nuclei are sufficiently accurate,
while those away from them are not accurate.
Hence, the inclusion of additional diffuse functions
is required for reliable computations, as shown in the previous subsection.
For the results shown in Table \ref{table:AAChirality},
several tens of percent error remain
between the triple and quadruple diffuse functions of alanine and valine.
Within this error, the results are considered to be settled.
The values are much smaller than those of H$_2$O$_2$,
and this difference is considered to originate in complicated structures of amino acid molecules.
In is noteworthy that 
all L-configuration amino acid molecules have positive electron chirality,
where the right-handed electron is the major component.
This means that 
D-amino acid molecules have a larger rate of the weak interaction
because the left-handed electron has a larger coupling constant of the weak interaction 
than the right-handed one.
This characteristic of L-amino acid is 
consistent with our scenario.

\begin{table*}[p]
  \centering
  \caption{Each orbital contribution to the total electron chirality of
alanine, serine, and valine
for the result of ONC+q-dyall.ae3z.
$E_i$ is the orbital energy,
and $E_{{\rm PV},i}$
is the orbital contribution to the parity-violating energy difference.
The value of the orbital contribution to electron chirality is shown for one electron in Kramers pair,
while both values in pairs are summed for the total.
}
  \begin{tabular}{c|rr||c|rr||c|rr} \hline\hline
\multicolumn{3}{c||}{Alanine } & \multicolumn{3}{c||}{Serine } & \multicolumn{3}{c}{Valine } \\
\hline 
No.   & $E_i$ [a.u.] & $\int d^3 x \psi_i^\dagger  \gamma_{5} \psi_i$ ~ & 
No.   & $E_i$ [a.u.] & $\int d^3 x \psi_i^\dagger  \gamma_{5} \psi_i$ ~ & 
No.   & $E_i$ [a.u.] & $\int d^3 x \psi_i^\dagger  \gamma_{5} \psi_i$ ~  
  \\ \hline \hline
    1     & ~ $-20.611$    & ~ $2.765\times10^{-10}$  & 
    1     & ~ $-20.615$    & ~ $4.012\times10^{-10}$  &
    1     & ~ $-20.609$    & ~ $9.202\times10^{-11}$   \\
    2     & $-20.548$    & $4.327\times10^{-11}$  &
    2     & $-20.591$    & $-4.221\times10^{-11}$  & 
    2     & $-20.546$    & $5.260\times10^{-11}$ \\
    3     & $-15.586$     & $-6.525\times10^{-11}$  &  
    3     & $-20.554$     & $5.745\times10^{-11}$  &  
    3     & $-15.583$     & $-8.182\times10^{-11}$ \\
    4     & $-11.374$     & $ 1.869\times10^{-10}$ & 
    4     & $-15.582$     & $-1.325\times10^{-10}$ & 
    4     & $-11.372$     & $ 6.424\times10^{-11}$  \\
    5     & $-11.295$     & $-6.265\times10^{-10}$ & 
    5     & $-11.379$     &  $3.640\times10^{-10}$ &  
    5     & $-11.289$     & $-5.354\times10^{-10}$ \\
    6     & $-11.236$     & $ 4.310\times10^{-10}$ & 
    6     & $-11.296$     & $ 1.311\times10^{-10}$ & 
    6     & $-11.246$     & $ 5.091\times10^{-10}$ \\
    7     & $-1.467$     & $-2.010\times10^{-9}$ &  
    7     & $-11.292$     & $-6.633\times10^{-10}$ &  
    7     & $-11.231$     & $-6.871\times10^{-11}$ \\
    8     & $-1.368$     & $-5.280\times10^{-10}$ &
    8     & $-1.472$     & $-2.617\times10^{-9}$ &
    8     & $-11.227$     & $ 8.182\times10^{-11}$  \\
    9     & $-1.226$     & $-1.340\times10^{-9}$  & 
    9     & $-1.391$     & $-1.475\times10^{-9}$  & 
    9     & $-1.465$     & $-7.300\times10^{-10}$ \\
   10     & $-1.029$     & $ 6.134\times10^{-9}$ & 
   10     & $-1.373$     & $-1.101\times10^{-9}$ & 
   10     & $-1.366$     & $-4.166\times10^{-10}$ \\
   11     & ~ $-0.905$    & $1.181\times10^{-8}$  &
   11     & ~ $-1.223$    & $-2.049\times10^{-9}$  &
   11     & ~ $-1.225$    & $-8.135\times10^{-10}$  \\
   12     & $-0.820$    & $5.880\times10^{-9}$ & 
   12     & $-1.022$    & $1.186\times10^{-8}$ & 
   12     & $-1.084$    & $-9.482\times10^{-10}$ \\
   13     & $-0.746$     & $-2.108\times10^{-8}$ &
   13     & $-0.897$     & $6.255\times10^{-9}$ &
   13     & $-0.964$     & $ 1.464\times10^{-8}$ \\
   14     & $-0.689$     & $ 3.974\times10^{-8}$  &
   14     & $-0.823$     & $ 4.068\times10^{-9}$  &
   14     & $-0.948$     & $-7.328\times10^{-10}$  \\
   15     & $-0.665$     & $-6.389\times10^{-8}$  & 
   15     & $-0.764$     & $5.669\times10^{-9}$  & 
   15     & $-0.832$     & $-1.056\times10^{-9}$  \\
   16     & $-0.637$     & $-1.641\times10^{-7}$  & 
   16     & $-0.711$     & $-4.008\times10^{-8}$  & 
   16     & $-0.795$     & $ 1.349\times10^{-8}$  \\
   17     & $-0.618$     & $ 2.808\times10^{-7}$  & 
   17     & $-0.685$     & $ 1.703\times10^{-7}$  & 
   17     & $-0.738$     & $-1.409\times10^{-8}$  \\
   18     & $-0.594$     & $-2.095\times10^{-8}$  & 
   18     & $-0.672$     & $ -3.330\times10^{-7}$  & 
   18     & $-0.697$     & $ 2.686\times10^{-8}$  \\
   19     & $-0.574$     & $-9.080\times10^{-8}$  & 
   19     & $-0.653$     & $ 1.712\times10^{-7}$  & 
   19     & $-0.668$     & $-7.062\times10^{-8}$ \\
   20     & $-0.534$     & $-1.516\times10^{-8}$  & 
   20     & $-0.631$     & $ 2.105\times10^{-7}$  & 
   20     & $-0.635$     & $-3.599\times10^{-9}$  \\
   21     & ~ $-0.527$    & $5.686\times10^{-8}$  &
   21     & ~ $-0.623$    & $-2.672\times10^{-7}$  &
   21     & ~ $-0.629$    & $-9.597\times10^{-8}$ \\
   22     & $-0.472$    & $-6.610\times10^{-8}$ &  
   22     & $-0.569$    & $ 6.071\times10^{-7}$ &  
   22     & $-0.619$    & $ 1.053\times10^{-7}$ \\
   23     & $-0.459$     & $ 4.793\times10^{-9}$ & 
   23     & $-0.559$     & $-4.959\times10^{-7}$ & 
   23     & $-0.588$     & $ 1.665\times10^{-7}$ \\
   24     & $-0.421$     & $ 3.971\times10^{-8}$  &
   24     & $-0.530$     & $-1.551\times10^{-7}$  &
   24     & $-0.582$     & $-1.060\times10^{-7}$  \\
        & & &
   25     & $-0.477$     & $ 5.944\times10^{-8}$  &
   25     & $-0.550$     & $ 5.656\times10^{-9}$   \\
        & & &
   26     & $-0.440$     & $ 7.650\times10^{-8}$  &
   26     & $-0.538$     & $ 6.256\times10^{-9}$  \\
        & & &
   27     & $-0.457$     & $-4.112\times10^{-8}$  &
   27     & $-0.509$     & $-1.138\times10^{-7}$  \\
        & & &
   28     & $-0.425$     & $ 1.554\times10^{-8}$  &
   28     & $-0.490$     & $ 8.997\times10^{-8}$  \\
& & &  & & &
   29     & $-0.487$     & $1.626\times10^{-8}$  \\
& & &  & & &
   30     & $-0.470$     & $-1.371\times10^{-7}$  \\
& & &  & & &
   31     & $-0.454$     & $8.728\times10^{-8}$  \\
& & &  & & &
   32     & $-0.413$     & $1.359\times10^{-8}$  \\
 \hline
    Total & & $ 2.954\times10^{-11}$ & & &$ 2.181\times10^{-10}$  & & & $ 2.952\times10^{-11}$
 \\ \hline
  \end{tabular}
  \label{Table:alanine_orbital}
\end{table*}

In Table \ref{Table:alanine_orbital},
the contribution from each orbital to the total electron chirality is shown.
The total electron chirality of all amino acid molecules 
was derived by the significant cancellation by two or three orders.
This indicates that the total electron chirality is strongly enhanced for excited or ionized states.

\section{Summary}

The total electron chirality in chiral amino acid molecules has been evaluated.
Alanine, serine, and valine were chosen for these molecules.
Previously, the total electron chirality of these light element molecules 
cannot be calculated 
owing to the smallness of the spin-orbit interaction.
We speculated that additional diffuse functions,
which are Gaussian functions with a small exponent,
are effective for this computation.
This speculation was based on the fact
that wave functions around nuclei are sufficiently accurate
and those away from them are not
because the results of the parity-violating energy difference are well-converged.
In this study, we confirmed our speculation
for the H$_2$O$_2$ molecule by numerical computation.
The results with additional diffuse functions are well converged 
and the dependence of the total electron chirality on the dihedral angle
is clearly consistent with our expectation.
Thus, we established the computational method
of the total electron chirality of molecules consisting of only light elements.
In this study, the total electron chirality of alanine, serine, and valine were evaluated
with additional diffuse functions.
It was determined that the values of the total electron chirality of L-alanine, L-serine, and L-valine
were much smaller than that of H$_2$O$_2$.
The total electron chirality of all L-amino acid molecules was positive.
This means that the number of right-handed electrons is larger than that of the left-handed ones
in L-amino acid molecules,
and D-amino acid molecules have a larger number of left-handed electrons
than right-handed ones.
Left-handed electron has a larger coupling constant of the weak interaction
and, hence, D-amino acid molecules exhibit a larger reaction rate of the weak interaction
than L-configuration.
Thus, D-amino acid molecules may be more broken in the evolution of the universe
by reactions with cosmic rays or neutrinos in space.
Therefore, ECCM may be the solution for the problem of homochirality in nature.
In addition, we showed the contribution from each orbital to the total electron chirality
for these amino acids.
We determined that the small total electron chirality is derived
by the cancellation between large contributions from HOMOs.
This indicates that the total electron chirality in excited and ionized states 
is much larger than that of the ground state,
as shown for imaginary doubly ionized states in our previous work \cite{Senami:2019}.
In our next study, 
we will confirm the enhancement of the total electron chirality in excited and ionized states.
For this confirmation, computations with the configuration interaction method will be performed.


\begin{acknowledgments}
This work was supported by Grants-in-Aid for Scientific Research (17K04982 and 19H05103).
\end{acknowledgments}

\end{document}